\documentclass[aps,prl,
twocolumn,
showpacs,nofootinbib, preprintnumbers,groupedaddress]{revtex4}
\usepackage{amssymb}
\usepackage{setspace}
\usepackage{amsmath}
\usepackage{latexsym}
\usepackage{graphicx}
\usepackage{graphics}

\begin{document}

\title{Elasticity and Viscosity of a Lyotropic Chromonic Nematic Studied
with Dynamic Light Scattering.}
\author{Yu. A. Nastishin$^{1,2}$}
\author{K. Neupane$^{3}$}
\author{A. R. Baldwin$^{3}$}
\author{O. D. Lavrentovich$^{1,4}$}
\email{odl@lci.kent.edu}
\author{S. Sprunt$^{3}$}
\email{ssprunt@kent.edu}
\affiliation{$^{1}$ Liquid Crystal Institute, Kent State University, POB 5190, Kent, OH
44242\\
$^{2}$ Institute of Physical Optics, 23 Dragomanov str., Lviv, 79005, Ukraine%
\\
$^{3}$ Department of Physics, Kent State University, Kent, OH, 44242\\
$^{4}$Chemical Physics Interdisciplinary Program, Kent State University,
Kent, OH 44242}
\date{\today}

\begin{abstract}

Using dynamic light scattering, we measure for the first time the
temperature-dependent elastic moduli and associated orientational viscosity
coefficients of the nematic phase in a self-assembled lyotropic chromonic
liquid crystal. The bend $K_{3}$ and splay $K_{1}$ moduli are an order of
magnitude higher than the twist $K_{2}$ constant. The ratio $K_{3}/K_{1}$
shows an anomalous increase with temperature; we attribute this to the
shortening of the aggregates as temperature increases. The viscosity
coefficients also show a significant anisotropy, as well as a strong
temperature dependence; in particular, the bend viscosity is three orders of
magnitude smaller than the splay and twist viscosities.
\end{abstract}


\maketitle

\newpage

Molecular self-assembly in solutions often results in anisometric aggregates
capable of orientational order. The simplest examples are end-to-end
\textquotedblleft living polymerization", formation of worm micelles by
surfactants, and face-to-face chromonic assembly of disc-like molecules into
stacks \cite{Lydon,Cates,Meyer_review}. In many systems, ranging from
organic dyes and drugs \cite{Lydon} to DNA \cite{Clark}, this mostly
one-dimensional aggregation produces a broadly polydisperse system of
\textquotedblleft building units" that form nematic or columnar phases and
are generally classified as lyotropic chromonic liquid crystals (LCLCs). The
structural organization and properties of LCLCs should depend on the
properties of both the individual molecules and their aggregates. Since the
length of the aggregates is not fixed by covalent bonds, it is expected to
vary strongly with concentration, temperature, ionic content, etc., making
the LCLCs very different from the classic liquid crystals in which the
building units are individual molecules of fixed shape. Despite the growing
interest in the LCLCs \cite{Lydon,Clark}, very little is known about their
molecular structure and practically nothing is known about their elastic and
viscous properties.

In this paper, we use dynamic light scattering from orientational
fluctuations to determine experimentally the fundamental elastic and
viscous properties of a nematic LCLC, following the approach
developed by Meyer's group for lyotropic polymer nematics
\cite{Meyer}. In particular, we study 14wt\% solutions of Disodium
Cromoglycate (DSCG) purchased from Spectrum Chemicals, 98\% purity,
in deionized water (initial resistivity 18 $M\Omega cm$). The
homogeneous nematic phase is stable up to the temperature
$T_{NI}\approx 301$ K, above which the mixture is in the biphasic
nematic - isotropic state. Our experiment was facilitated by the
recent development of aligning techniques for LCLCs
\cite{Opt_Charact} - specifically, the flat glass substrates used to
sandwich $16\mu m$ thick samples were coated with a buffed layer of
the SE-7511 polymer (Nissan Chemical Inc.) for uniform planar
orientation of the director $\vec{n}$ \cite{Opt_Charact}. The cells
were carefully sealed with epoxy to prevent water evaporation. To
mitigate possible effects of sample aging, we allowed the cells to
equilibrate for 48 hours; subsequent measurements were completed
within 15 hours. Longer equilibration worsens alignment, causing
parasitic static scattering. The sample temperature was controlled
within $0.1^{\circ }C$ by a hot stage (Instec). Light of wavelength
$\lambda =532$ nm from a diode-pumped solid state laser (Coherent,
model Verdi V8) was focused (beam waist $50\mu $m) normally onto the
sample. The incident light polarization was vertical; the analyzer
direction and the scattering plane were both horizontal.

\begin{figure}[tbp]
\begin{center}
\includegraphics[width=0.5 \textwidth]{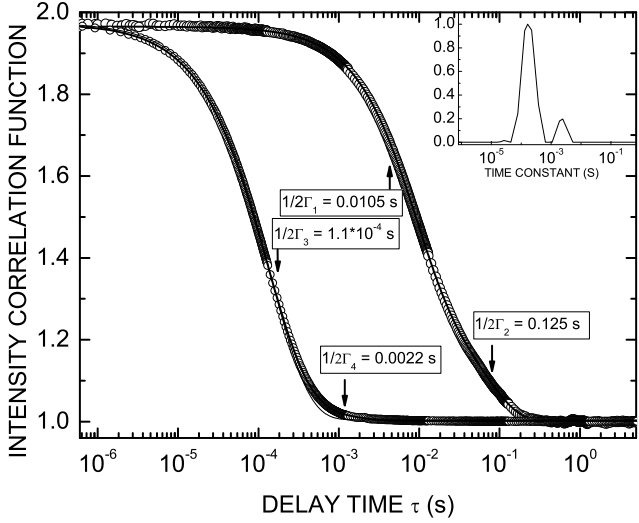}\\[0pt]
\end{center}
\caption{Correlation functions (open circles) collected at $T =294$ C for $q
= 1.64 \times 10^6 m^{-1}$ in a predominantly bend (left) and for $q=1.00
\times 10^7 m^{-1}$ in splay + twist geometries (right). Double-exponential
fits (thick lines) of the correlation functions give respectively: amplitude
and relaxation rate of splay $[A_{1},\Gamma _{1}]$, twist $[A_{2},\Gamma
_{2}]$, bend $[A_{3},\Gamma _{3}]$, and an additional $[A_{4},\Gamma _{4}]$
mode. The thin line for the bend correlation function (left) represents a
best single-exponential fit, which is clearly inadequate. Inset shows the
relaxation time spectrum obtained by the regularization method \protect\cite%
{regularization} for the bend geometry correlation function; the small
secondary peak confirms the presence of the additional mode $[A_{4},\Gamma
_{4}]$.}
\label{Correlationfunctions}
\end{figure}

\begin{figure}[tbp]
\begin{center}
\includegraphics[width=0.5 \textwidth]{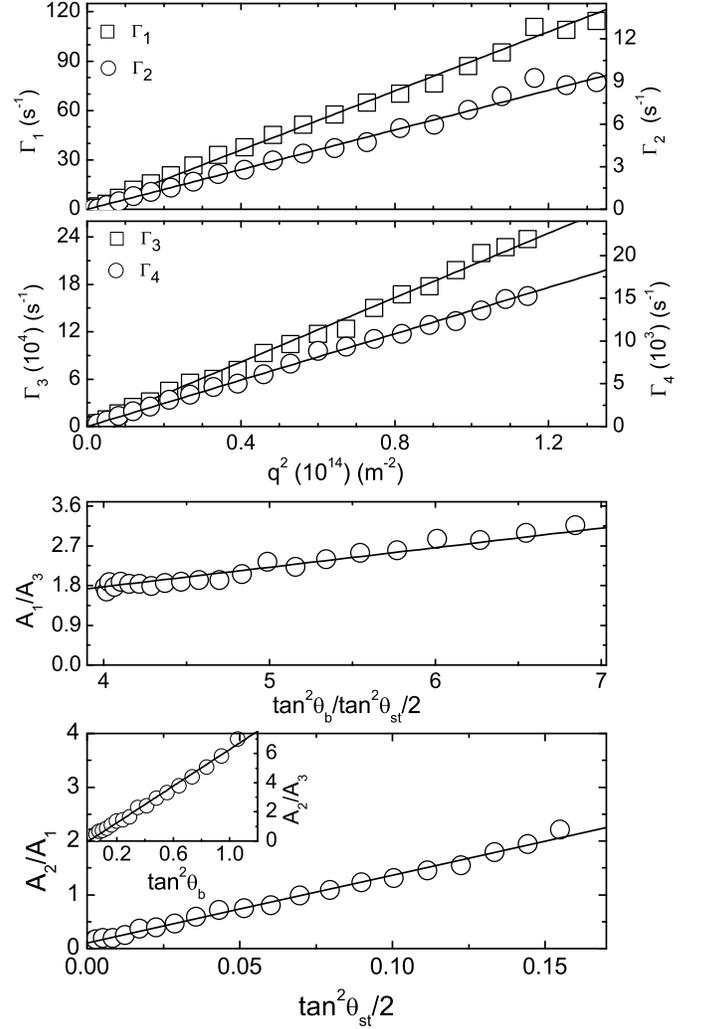}\\[0pt]
\end{center}
\caption{Angular dependencies of the rates and ratios of the amplitudes of
the relaxation modes. Solid lines represent fits described in the text.}
\label{RelaxratesAmps}
\end{figure}

In the two studied geometries the nematic director $\vec{n}$ is either (1)
perpendicular (splay + twist geometry) or (2) parallel (bend-twist geometry)
to the scattering plane. Homodyne cross-correlation functions of the
scattered light intensity (evenly split between two independent detectors)
were recorded as a function of time on a nanosecond digital correlator. In
geometry (1), two overdamped relaxation modes: splay [amplitude $A_{1}$,
relaxation rate $\Gamma_{1}]$ and twist $[A_{2},\Gamma_{2}]$ are expected
\cite{deGennes}. As determined by standard selection rules for nematic light
scattering, the faster mode corresponds to the splay director deformation,
and the slower mode assigns to the twist deformation. Typical time
correlation data are shown on the right side in Fig.\ref%
{Correlationfunctions}. Double exponential fitting of the decay of
bend-twist fluctuations for geometry (2), displayed on the left in Fig.\ref%
{Correlationfunctions}, combined with an alternative analysis by the
regularization method (see figure inset) \cite{regularization}, results in
an additional two modes with distinctive relaxation rates $\Gamma_{3}$ and $%
\Gamma_{4}$ and corresponding amplitudes $A_{3}$ and $A_{4}$. The best
single-exponential fit, shown by the thin line in Fig.\ref%
{Correlationfunctions}, does not match the slower portion of the decay. The
faster, more intense mode $[A_{3},\Gamma_{3}]$ is assigned to bend-twist
fluctuations, which are expected to dominate the depolarized scattering when
$\vec{n}$ lies in the scattering plane. The weaker mode $[A_4,\Gamma_4]$
cannot be attributed to a simple leakage of splay or twist fluctuations
potentially due to slight errors in setting geometry (2), as $\Gamma_4$
substantially exceeds both $\Gamma_1$ and $\Gamma_2$, even when the latter
two are measured at an order of magnitude larger wavenumber $q$. In this
context, we observe from Fig. \ref{RelaxratesAmps} that all detected modes
are hydrodynamic ($\Gamma \sim q^2$), as is normally expected for
orientational fluctuations.

For normally incident light, the amplitude and relaxation rate of the
bend-twist mode scale as $A_{3}\sim (K_{2}q_{\bot }^{2}+K_{3}q_{\Vert
}^{2})^{-1}$ and $\Gamma _{3}\sim K_{2}q_{\bot }^{2}+K_{3}q_{\Vert }^{2}$
\cite{deGennes}. (Here $\Vert $ means along $\vec{n}$.) For small scattering
angles, one has $K_{2}q_{\bot }^{2}\ll K_{3}q_{\Vert }^{2}$. Assuming for
the moment $K_{2}\approx K_{3}$, the ratio of these two terms is $(\sin
\theta_{b})^{2}/4n^{2}$, which ranges from $\approx 0.01$ for the lowest
scattering angle ($\theta_b = 15^{\circ }$) to $0.1$ for the largest ($%
65^{\circ }$). Here $n$ is the refractive index of the solution. In fact, as
shown below, $K_{3}\approx(30-40)K_{2}$, and thus the contribution of $%
K_{2}q_{\bot }^{2}$ to the amplitude and relaxation rate of the mode $%
[A_{3},\Gamma_{3}]$ is utterly negligible over the whole angular range
studied. Therefore, the dispersion of bend fluctuations can be accurately
measured while keeping the incident light normal to the sample substrates,
thus avoiding rotation of the sample in a $q$-scan and potential problems
associated with slight non-uniformity of alignment.

The variation of the relaxation rates and the amplitude ratios with
scattering wavenumber $q$ or scattering angle $\theta $ for the director
modes follows the expected behavior (solid lines in Fig. \ref{RelaxratesAmps}%
): $\Gamma _{i}=\frac{K_{i}}{\eta _{i}}q^{2}$; $\frac{A_{2}}{A_{1}}=\frac{%
K_{1}}{K_{2}}\tan ^{2}\frac{\theta _{st}}{2}$, $\frac{A_{2}}{A_{3}}=\frac{%
K_{3}}{4K_{2}}\tan ^{2}\theta _{b}$ and $\frac{A_{1}}{A_{3}}=\frac{K_{3}}{%
4K_{1}}\frac{\tan ^{2}\theta _{b}}{\tan ^{2}\frac{\theta _{st}}{2}}$, where $%
\eta _{i}$ with $i=1,2,3$ stand for viscosities corresponding to splay,
twist and bend deformations, respectively. Since the refractive index
anisotropy $\Delta n=0.02$ is small \cite{Opt_Charact}, one has $q=\frac{%
4\pi }{\lambda }n\sin \frac{\theta _{st}}{2}$ or $q=\frac{2\pi }{\lambda }%
n\sin \theta _{b}$, where $\theta _{st}$ and $\theta _{b}$ are the
scattering angles, in the splay+twist and bend geometries, respectively.
Good fitting of the relaxation rates and the amplitudes indicates that data
collected during the experiment are not significantly affected by sample
aging.

\begin{figure}[tbp]
\begin{center}
\includegraphics[width=0.50 \textwidth]{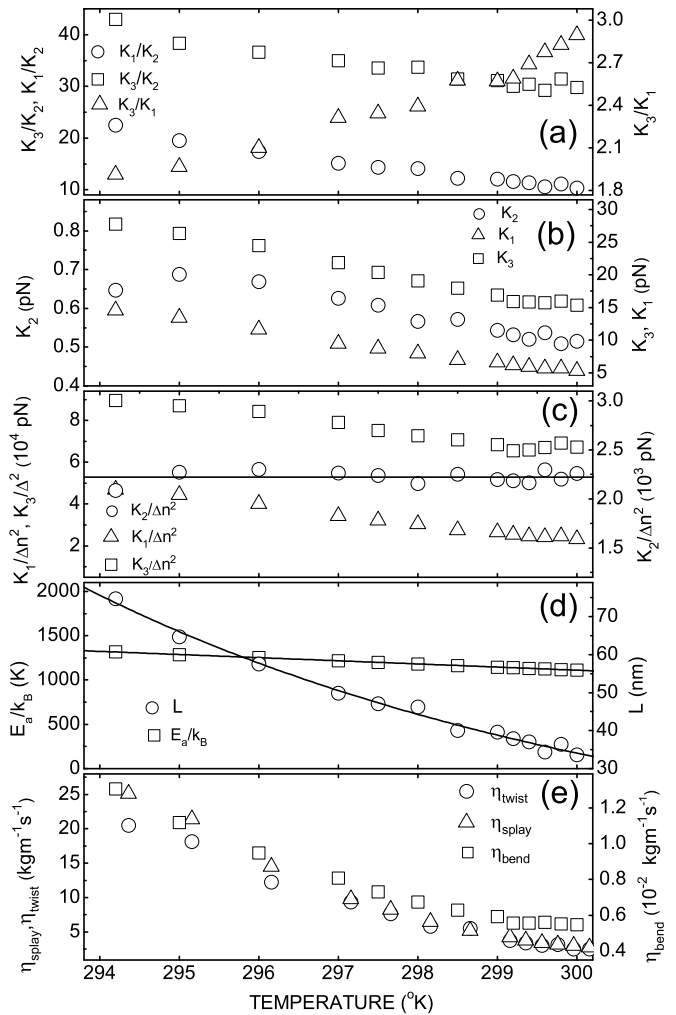}\\[0pt]
\end{center}
\caption{Temperature dependence of the elastic moduli (a, b), reduced moduli
(c), parameters $L$ and $E_{a}$ (d), and orientational viscosities (e), for
the studied chromo-nematic.}
\label{ElasticViscosity}
\end{figure}

From data for the amplitudes measured on heating at fixed scattering angles (%
$\theta _{st}=40^{\circ }$ for splay+twist and $\theta _{b}=15^{\circ }$ for
bend geometries, respectively) we deduced the temperature dependencies of
the elastic moduli ratios (Fig. \ref{ElasticViscosity} a). Absolute values
of the moduli (Fig. \ref{ElasticViscosity} b) were determined by calibrating
the scattered intensity against an identical cell filled with the standard
thermotropic nematic (4'-n-pentyl-4-cyanobiphenyl, 5CB) for a pure bend
geometry at 294 K. We then used the known $K_{3}=18$ pN, $\Delta n=0.201$
for 5CB \cite{KaratMadhusudana}, and the temperature dependence of $\Delta n$
for our chromonematic \cite{Opt_Charact}, to obtain $K_{3}=28$ pN at 294 K.
Finally we combined the results for the elasticities with data for the
relaxation rates to find the associated orientational viscosities (Fig. \ref%
{ElasticViscosity}e).

Within the phenomenological Landau-de Gennes theory, it is expected that the
temperature behavior of $K_{i}\sim S^{2}$ and $\Delta n\sim S$ are
determined by the temperature dependency of the scalar order parameter $S$
\cite{deGennes}. In the chromonematic case, we find that only the twist mode
follows this expectation, as the amplitude $A_{2}\sim \frac{(\Delta n)^{2}}{%
K_{2}}$ is temperature independent \cite{deGennes} (Fig. \ref%
{ElasticViscosity} c). In contrast, $K_{3}/(\Delta n)^{2}$ and $%
K_{1}/(\Delta n)^{2}$ both decrease as temperature increases. The effect can
be qualitatively explained by shortening of aggregates with increasing $T$
\cite{LS_Isotrop}, which should affect $K_{1}$ and $K_{3}$, but not
significantly $K_{2}$ \cite{DeGennes_review}. Meyer \cite{Meyer_review}
predicted that in a self-assembled chromonematic, $K_{1}=\frac{k_{B}T}{4d}%
\frac{L}{d}$, where $L=L_{0}\exp \frac{E_{a}}{2k_{B}T}$ is the average
aggregate length \cite{Cates}, $L_{0}$ is a constant, $E_{a}$ is the
scission energy needed to split an aggregate into two, $k_{B}$ is the
Boltzmann constant, and $d$ is the diameter of the stack. Qualitatively,
Meyer's model describes the data well (Fig. \ref{ElasticViscosity} d). It is
of interest to explore whether the fitting parameters are close to the
physical expectations.

For the analysis, we consider $K_{1}/K_{2}=\frac{1}{K_{2}^{0}}\frac{k_{B}T}{%
4d}\frac{L}{d}$ (to avoid the possible role of $S$) where the coefficient $%
K_{2}^{0}$ is taken as the value of $K_{2}/S^{2}$ at $T=294$ K with the
experimentally determined $K_{2}=0.65$ pN and the typical $S=0.7$ \cite%
{Opt_Charact,Collings}. Then with $K_{2}^{0}=1.32$ pN and $d=1.6$ nm \cite%
{HartshorneWoodard}, we use the experimental data for $K_{1}/K_{2}$ to
deduce $L_{0}=0.86$ nm, $L(T)$, and $E_{a}$ (as shown in Fig. \ref%
{ElasticViscosity} d). $L$ decreases from $75$ nm at $294\ $K to $34$ nm at $%
300$ K \cite{remark}, which is consistent with the estimate $L=18$ nm \cite%
{LS_Isotrop} for the isotropic phase at 305K. Furthermore, the scission
energy shows a weak temperature dependency, changing from $E_{a}=3.3\times
10^{-20}J$ at 294K to $E_{a}=2.8\times 10^{-20}J$ at 300K, which amounts to $%
(7-8)k_{B}T$, in agrement with estimates for chromonic dyes and DNA
oligomers \cite{Collings,Clark}.

The case of $K_{3}$ is the most difficult to discuss as it might depend not
only on $L$ but also on the persistence length $P$ of the aggregates. In the
model of rigid rods, $P\rightarrow \infty$, and $K_{3}/K_{1}$ $\varpropto
L/d $ should decrease with increasing $T$ if the aggregates become shorter
\cite{DeGennes_review}, contrary to the experimentally observed behavior in
our chromonematic (Fig. \ref{ElasticViscosity} a). The observed increase of $%
K_{3}/K_{1}$ with $T$ might be consistent with the model of the flexible
rods, in which \cite{DeGennes_review} $K_{3}\varpropto P$ is independent of $%
L$ while $K_{1}\varpropto L$; however, the data on $P$ are not presently
available.

The viscosity coefficients $\eta _{splay}$, $\eta _{twist}$ and $\eta _{bend}
$ all decrease on heating (Fig. \ref{ElasticViscosity}e), as in other
nematics. It is of interest to compare them to the viscosities measured (at
room temperatures) for the low-molecular weight 5CB \cite{Kelly}, and for
the lyotropic polymer nematic poly-$\gamma $-benzyl-glutamate (PBG) \cite%
{Meyer} : $\eta _{bend}$ (5CB : PBG : DSCG) = (0.028 : 0.016 : 0.013) kgm$%
^{-1}$s$^{-1}$ and $\eta _{twist}$ (5CB : PBG : DSCG) = (0.08 : 3.5 : 20.5)
kgm$^{-1}$s$^{-1}$. Clearly, $\eta _{bend}$ in all three systems is of the
same order, while $\eta _{twist}$ in the polymer and chromonematic is
two-three orders of magnitude higher than in 5CB. The large ratio of $\eta
_{twist}/\eta _{bend}\approx 1600$ measured for DSCG is not unusual from the
point of view of the theory \cite{Meyer} and resemble findings for the
polymer nematics \cite{LeeMeyer}: $\eta _{bend}$ for sufficiently extended
aggregates is associated mainly with the relatively easy process of sliding
of the aggregates along each other, while $\eta _{twist}$ should involve
rearrangements of the aggregates. Interestingly, $\eta _{twist}/\eta _{bend}$
decreases dramatically with $T$, to $\approx 450$ at 300K. Such a behavior
can be tentatively explained by the fact that the aggregates become shorter
at high $T$, which should dramatically decrease $\eta _{twist}$ and less so $%
\eta _{bend}$. It is of interest to explore in the future the role of
possible breaking and recombination of chromonic aggregates in the dynamics
of LCLCs.

An interesting phenomenon observed in the course of our experiments
is the development at a threshold laser power of a strong forward
diffraction pattern. Evidently a transformation of the sample
structure takes place above the threshold power, which was about 30
mW at $T=294$ K, and decreased on heating, vanishing near the
temperature $T_{NI}=301$ K. To avoid any significant influence on
our results reported above, we kept the laser power as low as
feasible between 1.5 and 3 mW and well below the threshold for all
$T$ studied. This is also why we do not include in
Fig.\ref{ElasticViscosity} data measured in the vicinity of
$T_{NI}$, which might be affected by the laser power.

It remains to discuss the additional mode $[A_{4},\Gamma _{4}]$
observed in the bend geometry. Since its amplitude is $\sim 7$ times
smaller than the amplitude of the bend mode, its influence, if any,
on the results obtained in the bend geometry, is small: $K_{3}$
might be slightly overestimated (by less than $15\%$). Concerning
the origin of this mode, we can suggest that it corresponds to the
translation diffusion either of monomers, short aggregates ($L\sim
d$) or to a transverse sliding of DSCG
molecules or their clusters within the overall aggregates. From data for $%
\Gamma _{4}$ shown in Fig. \ref{RelaxratesAmps}b using Einstein's formula
for the diffusion coefficient of spherical scatterers, we find the
hydrodynamic diameter $d_{h}=3.2$ nm, which is twice the aggregate diameter $%
d\approx 1.6$ nm \cite{HartshorneWoodard} and could reflect short
aggregates. However the studied system is expected to be poly-disperse in
length and it is not likely that only the diffusion of short aggregates
would be detected. On the other hand, if we assume that the mode $%
[A_{4},\Gamma _{4}]$ corresponds to the lateral sliding, we have to account
for the drag force from the neighboring molecules. According to ref. \cite%
{LiTang} this can be done by replacing the solvent viscosity $\eta $ with
the effective viscosity $\eta _{eff}=c\eta $, where the drag coefficient $%
c\approx 2$ \cite{LiTang}. If we use $\eta _{eff}$ instead of $\eta $, we
arrive at $d_{h}\approx d$. It is thus plausible that the mode $%
[A_{4},\Gamma _{4}]$ corresponds to the sliding of molecules within
the aggregates. This sliding can provide a bending flexibility
mechanism for the aggregates that differs from the known bending
mechanisms in other rod-like nematics.

To conclude, we have measured for the first time the temperature
dependencies of the elastic moduli and corresponding viscosities for a new
type of lyotropic nematic, the self-assembled chromonic nematic. The
observed anomalous temperature behavior of the splay and bend elastic
constants indicates that the system is different from the well studied
nematic phases of thermotropic low-molecular weight liquid crystals and
their lyotropic counterparts formed either by hard rods or flexible
filaments. One of the key mechanisms contributing to these differences is
the temperature dependence of the average aggregate length. These and other
features increasingly point to the chromonematics as a phenomenologically
richer system than liquid crystals formed by building units of a fixed size.

This work was supported by NSF Materials World Network on Lyotropic
Chromonic Liquid Crystals, grant DMR076290, by Samsung Electronics
Corp. and by Fundamental Research State Fund Project UU24/018.



\begin{references}
\bibitem{remark} The data remain essentially the same when one uses
$\frac{K_1}{(\Delta n)^{2}}$ for analysis. If one uses $K_{1}(T)$
without $S$-normalization then the fitted $L$ varies from 39 nm to
13 nm in the same temperature interval.
\end{references}
\end{document}